\documentclass[sn-mathphys,Numbered]{sn-jnl}

\usepackage{graphicx}%
\usepackage{multirow}%
\usepackage{amsmath,amssymb,amsfonts}%
\usepackage{amsthm}%
\usepackage{mathrsfs}%
\usepackage[title]{appendix}%
\usepackage{xcolor}%
\usepackage{textcomp}%
\usepackage{manyfoot}%
\usepackage{booktabs}%
\usepackage{algorithm}%
\usepackage{algorithmicx}%
\usepackage{algpseudocode}%
\usepackage{listings}%
\usepackage{epsfig}

\begin{document}

\title[Asymptotic behaviour with anomalous diffusion]{Asymptotic behaviour for convection with anomalous diffusion}

\author[1]{\fnm{Brian} \sur{Straughan}}\email{brian.straughan@durham.ac.uk}

\equalcont{These authors contributed equally to this work.}

\author*[2]{\fnm{Antonio} \sur{Barletta}}\email{antonio.barletta@unibo.it}

\equalcont{These authors contributed equally to this work.}

\affil[1]{\orgdiv{Department of Mathematics}, \orgname{University of Durham}, \orgaddress{\street{Stockton Road},
\city{Durham}, \postcode{DH1 3LE}, \country{U.K.}}}

\affil[2]{\orgdiv{Department of Industrial Engineering}, \orgname{University of Bologna},
\orgaddress{\street{Viale Risorgimento 2}, \city{Bologna}, \postcode{40136}, \country{Italy}}}

\abstract{
We investigate the fully nonlinear model for convection in a Darcy porous material where the diffusion is of anomalous type
as recently proposed by Barletta.
The fully nonlinear model is analysed but we allow for variable gravity or penetrative convection effects which result in
spatially dependent coefficients. This spatial dependence usually requires numerical solution even in the linearized case.
In this work we demonstrate that regardless of the size of the Rayleigh number the perturbation solution will decay
exponentially in time for the superdiffusion case. In addition we establish a similar result for convection in a bidisperse 
porous medium where both macro and micro porosity effects are present. Moreover, we demonstrate a similar result for 
thermosolutal convection.
}





\maketitle

\section{Introduction}	\label{intro}
\setcounter{equation}{0}

Anomalous or fractional heat and mass diffusion defines deviations from the traditional diffusion model, which describes the movement of molecules as a random walk with a constant diffusion coefficient. In anomalous diffusion, the mean squared displacement of molecules grows with time slower or faster than linearly. With a slower growth, the regime is subdiffusive, meaning molecules spread out more slowly than expected. When the growth is faster than linear, the process is superdiffusive, meaning molecules spread out more rapidly than in standard diffusion. A thorough review on this topic can be found in \citet{henry2010introduction}. 
In recent work, \citet{Barletta:2023} has proposed a model for convective motion of a fluid in a saturated porous medium where
the diffusion coefficient is time dependent but may be of subdiffusion or superdiffusion type. He showed, for the linearized theory,
that regardless of the size of the Rayleigh number the solution to the perturbation system may grow in the short term, but eventually
the perturbation will decay for superdiffusion, and will grow indefinitely for subdiffusion.
The object of this paper is to analyse the completely nonlinear problem, and we show that exponential decay holds in this case
for superdiffusion. We establish this result in a saturated porous material but also allow for variable gravity effects,
cf. \citet{Straughan:1989}, or for penetrative convection driven by an internal heat source, cf. \citet[~p. 343]{Straughan:2004}.
This extension is important because mathematically the coefficients in the perturbation equations often depend on the vertical
coordinate $z$. For such spatial dependence it is usually not possible to obtain an analytical solution by a normal mode
procedure, cf. \citet{Barletta:2019,Barletta:2021}, and one must resort to numerical techniques. We additionally establish 
nonlinear decay of the solution when the porous medium is bidisperse, i.e. the medium has macro pores, but the solid skeleton
possesses fissures or cracks, necessitating the inclusion of micropores, and hence a bidisperse or double porosity structure.
Furthermore, we prove a similar asymptotic decay result when thermal and salt effects are present, i.e. in the case of 
thermosolutal or double diffusive convection 

The extension to spatially dependent coefficients is essential for real life applications. Also, consideration of
double diffusive flow in a bidisperse porous medium is a topic of immense importance in current everyday life. For example, 
this phenomenon is proving important in renewable energy research, especially in solar pond technology, see
\citet{DineshkumarRaja:2022}, \citet{Wang:2018}. A particularly important application of double diffusion in a bidisperse
porous medium is to magma flow in a volcano, see e.g. \citet{Vieira:2021}, \citet{Allocca:2022}, \citet{Toy:2019}, 
\citet{Singh:2022}, \citet{BagdassarovFradkov:1993}, \citet{DeCampos:2005}. Given the recent seismic and volcanic activity 
in the Campi Flegrei region near Pozzuoli, an understanding of this scenario, 
and the potential effects of anomalous diffusion, 
is of vital importance.

We now present the equations for convection models and deliver a fully nonlinear analysis of asymptotic solution behaviour.

\section{Thermal convection in a Darcy porous material} \label{equations}

The subject of thermal convection in a Darcy porous material is investigated by \citet{Barletta:2023} who allows for
anomalous thermal diffusion by introducing a statistically motivated time dependent diffusion coefficient leading to a
diffusion of form $\varphi D_rrt^{r-1}\Delta C$, where 
$C$
may be a concentration or a temperature field. The coefficients 
$\varphi,D_r,r$
are positive constants being porosity, constant diffusion coefficient, and an exponent. The variable
$t$
denotes time and
$\Delta$
is the three - dimensional Laplace operator
\begin{equation*}
\Delta=
\frac{\partial^2}{\partial x^2}
+\frac{\partial^2}{\partial y^2}
+\frac{\partial^2}{\partial z^2}\,.
\end{equation*}
The basic solution of \citet{Barletta:2023} is the classical one of \citet{Chandrasekhar:1981}, where
$C$
(or
$T$
in the case of temperature) is linear in the vertical coordinate
$z$
and the velocity is 0.

We commence with the non-dimensional perturbation equations of \citet[eqs. ~ (36) - (38)]{Barletta:2023}, which employ
a Boussinesq approximation, cf. \citet{Barletta:2022}, and have form (in our notation)
\begin{equation}\label{E:Pertn}
\begin{aligned}
&u_i+\pi_{,i}-  Ra\theta k_i=0,\\
&u_{i,i}=0,\\
&\theta_{,t}+u_i\theta_{,i} = w + k t^{r-1}\Delta \theta,
\end{aligned}
\end{equation}
where
$u_i,\pi,\theta$
are the velocity perturbation, pressure perturbation, and temperature perturbation,
$w=u_3$,
${\bf k}=(0,0,1)$,
and 
$Ra$
is the Rayleigh number. It is convenient to rescale these equations and employ the parameter
$R=\sqrt{Ra}$
and we write \eqref{E:Pertn} in the form
\begin{equation}\label{E:PertnN}
\begin{aligned}
&u_i=-\pi_{,i}+ R\theta k_i,\\
&u_{i,i}=0,\\
&\theta_{,t}+u_i\theta_{,i}=Rw+kt^{r-1}\Delta \theta.
\end{aligned}
\end{equation}
Throughout we investigate the superdiffusion case where
$r>1$.

\citet{Barletta:2023} essentially linearizes \eqref{E:PertnN} and employs a normal mode analysis, cf. 
\citet{Barletta:2019,Barletta:2021}, to derive very interesting novel behaviour. 
In the interests of encompassing greater physical 
behaviour we allow for the effect of variable gravity, cf. \citet{Straughan:1989}, and for
penetrative convection driven by an internal heat source, cf. \citet[p.~343]{Straughan:2004},
and the perturbation equations \eqref{E:PertnN} are replaced by
\begin{equation}\label{E:PertA}
\begin{aligned}
&u_i=-\pi_{,i}+ Rg(z)\theta k_i,\\
&u_{i,i}=0,\\
&\theta_{,t}+u_i\theta_{,i}=Rh(z)w+kt^{r-1}\Delta \theta,
\end{aligned}
\end{equation}
where
$g,h$
are bounded functions of
$z$, $k$
is a positive constant and
$r>1$. The domain for equations \eqref{E:PertA} is
$(x,y)\in\mathbb{R}^2$, $\{z\in(0,1)\}$
and
$t>0$.
The boundary conditions are
\begin{equation}\label{E:BCs}
w=0,\theta=0,\quad z=0,1,
\end{equation}
together with 
$u_i,\pi,\theta$
being periodic in
$x,y$.
The periodicity ensures cellular structure of the convection cells as explained in detail by 
\citet[pp.~43-52]{Chandrasekhar:1981}.

Suppose
\begin{equation}\label{E:Bds}
\vert g(z)\vert\le c_2,\qquad \vert g+h\vert \le c_1, \qquad \forall z\in [0,1],
\end{equation}
for constants 
$c_1$
and
$c_2$.
Then let
$V$
be a periodic cell for the perturbation solution to \eqref{E:PertA}. Let
$(\cdot, \cdot)$
and
$\Vert\cdot\Vert$
be the inner product and norm on
$L^2(V)$.

Multiply 
\eqref{E:PertA}$_1$
by
$u_i$
and integrate over
$V$, 
and multiply
\eqref{E:PertA}$_3$
by
$\theta$
and integrate over
$V$.
After integration by parts and use of the boundary conditions one may show
\begin{equation}\label{E:I1}
0=-\Vert{\bf u}\Vert^2+R(gw,\theta),
\end{equation}
and
\begin{equation}\label{E:I2}
\frac{d}{dt}\frac{1}{2}\Vert\theta\Vert^2=R(hw,\theta)-kt^{r-1}\Vert\nabla\theta\Vert^2.
\end{equation}
We next add \eqref{E:I1} and \eqref{E:I2} and then use the arithmetic - geometric mean inequality on the result to obtain
\begin{equation}\label{E:Ineq1}
\frac{d}{dt}\frac{1}{2}\Vert\theta\Vert^2\le \frac{R^2c_1^2}{4}\Vert\theta\Vert^2-kt^{r-1}\Vert\nabla\theta\Vert^2.
\end{equation}
The function
$\theta$
satisfies Poincar\'e's inequality
$\Vert\nabla\theta\Vert^2\ge\pi^2\Vert\theta\Vert^2$
and then from \eqref{E:Ineq1} we derive
\begin{equation}\label{E:Ineq2}
\frac{d}{dt}\Vert\theta\Vert^2+\Bigl(2kt^{r-1}\pi^2-\frac{R^2c_1^2}{2}\Bigr)\Vert\theta\Vert^2\le 0.
\end{equation}
Now employ an integrating factor and integrate \eqref{E:Ineq2} to see that
\begin{equation}\label{E:Ineq3}
\Vert\theta(t)\Vert^2\le\Vert\theta(0)\Vert^2\,\exp\Bigl(\frac{R^2c_1^2t}{2}-\frac{2k\pi^2}{r}t^r\Bigr).
\end{equation}
It follows from \eqref{E:Ineq3} that as
$t\to\infty$,
$\Vert\theta(t)\Vert$
tends to zero very rapidly no matter how large 
$\Vert\theta(0)\Vert$
is.

From \eqref{E:I1} one shows
\begin{equation*}
\Vert{\bf u}\Vert^2\le R^2c_1^2\Vert\theta\Vert^2,
\end{equation*}
and so decay of
$\Vert\theta(t)\Vert$
also implies decay of
$\Vert{\bf u}(t)\Vert$.

\vskip12pt

\noindent{\bf Remark}

\noindent The exponential decay is obtained regardless of the size of the Rayleigh number 
$Ra=R^2$.
The result \eqref{E:Ineq3} applies also to the Barletta model where 
$g=h=1$. 
Observe that we here employ no linearization and show decay for a solution to the fully nonlinear equations.

\section{Convection in a bidisperse porous material}

For a single temperature,
$T$,
in the macro and micro pores, equations for thermal convection in a bidisperse Darcy porous material are given by
\citet{GentileStraughan:2017}. If we adopt the anomalous diffusion term of \citet{Barletta:2023}, then the non-dimensional
perturbation equations for thermal convection in a bidisperse porous medium may be shown to be, cf.
\citet{GentileStraughan:2017}, \citet{Straughan:2018prsa},
\begin{equation}\label{E:Bidis1}
\begin{aligned} 
&u_i^f+\xi(u_i^f-u_i^p)=-\pi^f_{,i}+R\theta k_i,\\
&u^f_{i,i}=0,\\
&K_ru_i^p+\xi(u_i^p-u_i^f)=-\pi^p_{,i}+R\theta k_i,\\
&u^p_{i,i}=0,\\
&\theta_{,t}+(u^f_i+u^p_i)\theta_{,i}=R(w^f+w^p)+kt^{r-1}\Delta\theta.
\end{aligned} 
\end{equation} 
Here
$u^f_i,u^p_i,\pi^f,\pi^p$
and
$\theta$
are nonlinear perturbations to the velocity in the macropores, velocity in the micropores, 
pressure in the macropores,
pressure in the micropores,
and temperature, respectively, with
$w^f=u^f_3,w^p=u^p_3.$
The parameters
$\xi$
and
$K_r$
are an interaction coefficient and the relative permeability
$K_r=K^f/K^p$,
where
$K^f$
and 
$K^p$
are the permeabilities in the macro and micro pores.

Equations \eqref{E:Bidis1} hold on
$\{(x,y)\in\mathbb{R}^2\},$
$\{z\in(0,1)\}, t>0$,
and the boundary conditions are
\begin{equation}\label{E:BidisBC}
w^f=0, w^p=0, \theta=0,\qquad{\rm on}\quad z=0,1,
\end{equation}
together with 
$u_i^f,u_i^p,\pi^f,\pi^p,\theta$
being periodic in
$x,y$.

To derive an asymptotic behaviour result we multiply \eqref{E:Bidis1}$_1$ by
$u_i^f$,
\eqref{E:Bidis1}$_3$ by
$u_i^p$
and \eqref{E:Bidis1}$_5$ by
$\theta$
and integrate each over 
$V$
using integration by parts and the boundary conditions. After addition of the equations for
$u_i^f$
and
$u_i^p$
this leads to
\begin{equation}\label{E:I5}
0=-\Vert{\bf u}^f\Vert^2-K_r\Vert{\bf u}^p\Vert^2-\xi\Vert{\bf u}^f-{\bf u}^p\Vert^2
+R(\theta,w^f+w^p),
\end{equation}
and
\begin{equation}\label{E:I6}
\frac{d}{dt}\frac{1}{2}\Vert\theta\Vert^2=
R(\theta,w^f+w^p)
-kt^{r-1}\Vert\nabla\theta\Vert^2.
\end{equation}
We now add equations \eqref{E:I5} and \eqref{E:I6} and we use the arithmetic - geometric mean inequality on the
$(\theta,w^f+w^p)$
terms to arrive at
\begin{equation}\label{E:I7}
\frac{d}{dt}\frac{1}{2}\Vert\theta\Vert^2\le
R^2(1+K_r)\Vert\theta\Vert^2
-kt^{r-1}\Vert\nabla\theta\Vert^2.
\end{equation}
Now employ Poincar\'e's inequality on the last term and use an integrating factor as before to obtain
\begin{equation}\label{E:I8}
\Vert\theta(t)\Vert^2\le
\Vert\theta(0)\Vert^2\Bigl(
2R^2(1+K_r)t-\frac{2k\pi^2}{r}t^r
\Bigr)\,.
\end{equation}
This establishes that since
$r>1$,
$\Vert\theta(t)\Vert$
decays rapidly as
$t\to\infty$,
for any
$Ra$
and 
$\Vert\theta(0)\Vert$.

From \eqref{E:I5} one shows
\begin{equation}\label{E:I9}
\Vert{\bf u}^f\Vert^2+K_r\Vert{\bf u}^p\Vert^2\le R^2\Bigl(1+\frac{1}{K_r}\Bigr)\Vert\theta\Vert^2.
\end{equation}
Hence, from \eqref{E:I8} decay of
$\Vert\theta(t)\Vert$
guarantees decay of both
$\Vert{\bf u}^f(t)\Vert$
and
$\Vert{\bf u}^p(t)\Vert$.

\section{Double diffusive porous convection}

In the case of double diffusive convection in a Darcy porous material when the layer is heated below and salted above or below,
the basic solution is as in e.g. \citet[pages~238-241]{Straughan:2004}, and the perturbation equations are given as,
\citet[eqs.~(14.20)-(14.22)]{Straughan:2004}. If we started at the outset with the anomalous diffusion of \citet{Barletta:2023} 
for the temperature of form
$k_1t^{r-1}\Delta\theta$
and for salt as
$k_2t^{s-1}\Delta\phi$,
for constants
$k_1,k_2$
and 
$r>1, s>1$,
where
$\phi$
is now the salt perturbation, then the non-dimensional perturbation equations are
\begin{equation}\label{E:DDeqs}
\begin{aligned}
&0=-u_i+R\theta k_i-R_s\phi k_i,\\
&u_{i,i}=0,\\
&\theta_{,t}+u_i\theta_{,i}=Rw+k_1t^{r-1}\Delta\theta,\\
&Le(\phi_{,t}+u_i\phi_{,i})=\pm R_sw+k_2t^{s-1}\Delta\phi,
\end{aligned}
\end{equation}
where
$Ra=R^2$
is the Rayleigh number,
$\mathcal{C}=R_s^2$
is the salt Rayleigh number and
$Le$
is a constant known as the Lewis number.

We allow for different exponents of anomalous diffusion and without loss of generality we here assume
$s>r>1$.
We consider the case of the minus sign in \eqref{E:DDeqs}$_4$
which corresponds to salting from above. The analysis in the other case of the plus sign is easier and we omit details.

To derive an asymptotic result we note that for
$t\in(0,1), t^{r-1}>t^{s-1}$.
From \eqref{E:DDeqs}$_1$
we may obtain
\begin{equation}\label{E:D1}
0=-\Vert{\bf u}\Vert^2+R(\theta,w)-R_s(\theta,w).
\end{equation}
From \eqref{E:DDeqs}$_3$ and \eqref{E:DDeqs}$_4$
one finds
\begin{equation}\label{E:D2}
\begin{aligned}
\frac{d}{dt}\Bigl(
\frac{1}{2}\Vert\theta\Vert^2
+\frac{Le}{2}\Vert\phi\Vert^2
\Bigr)
=&R(w,\theta)-k_1t^{r-1}\Vert\nabla\theta\Vert^2\\
&+R_s(w,\phi)-k_2t^{s-1}\Vert\nabla\phi\Vert^2.
\end{aligned}
\end{equation}
We add \eqref{E:D1} and \eqref{E:D2} and use the arithmetic - geometric mean inequality and Poincar\'e's inequality to
obtain
\begin{align}
\frac{d}{dt}\Bigl(
\frac{1}{2}\Vert\theta\Vert^2
+\frac{Le}{2}\Vert\phi\Vert^2
\Bigr)
\le&
(2R^2-k_1t^{r-1}\pi^2)\Vert\theta\Vert^2
+(2R^2_s-k_2t^{s-1}\pi^2)\Vert\phi\Vert^2\notag\\
\le&
(2R^2-k_1t^{s-1}\pi^2)\Vert\theta\Vert^2
+(2R^2_s-k_2t^{s-1}\pi^2)\Vert\phi\Vert^2.\label{E:D3}
\end{align}
Now put
$A=\max\{4R^2,4R_s^2\},$ $B=\min\{2k_1,2k_2\}$,
and from \eqref{E:D3} we may obtain
\begin{equation}\label{E:D4}
\frac{dF}{dt}+(B\pi^2t^{s-1}-A)F\le 0,
\end{equation}
for
$t\in(0,1]$,
where
$F(t)=\Vert\theta\Vert^2+Le\Vert\phi\Vert^2$.
Integrate \eqref{E:D4} with an intergrating factor to obtain
\begin{equation*}
F(1)\exp \Bigl(\frac{B\pi^2}{s}-A\Bigr)\le F(0)e^{-A}.
\end{equation*}

We now apply a similar argument to the above on the time interval
$(1,\infty)$,
observing then that
$t^{s-1}>t^{r-1}>1.$
In this case we obtain instead of \eqref{E:D4},
\begin{equation}\label{E:D5}
\frac{dF}{dt}+(B\pi^2t^{r-1}-A)F\le 0,
\end{equation}
This is integrated with an integrating factor to arrive at
\begin{align*}
F(t)\exp\Bigl(\frac{B\pi^2}{r}t^r-At\Bigr)\le& F(1)\exp \Bigl(\frac{B\pi^2}{r}-A\Bigr)\\
\le&F(0)\exp\Bigl[\frac{B\pi^2(r-s)}{rs}-A\Bigr]=G(0),
\end{align*}
where
$G(0)$
is as defined. Thus,
\begin{equation}\label{E:D6}
\Vert\theta(t)\Vert^2+Le\Vert\phi(t)\Vert^2\le G(0)\exp\Bigl(At-\frac{B\pi^2}{r}t^r\Bigr)\,.
\end{equation}
Inequality \eqref{E:D6} shows that as
$t\to\infty$,
$\Vert\theta(t)\Vert$
and
$\Vert\phi(t)\Vert$
both decay rapidly regardless of the size of
$Ra$, $\mathcal{C}$,
or the initial data.

By using the arithmetic - geometric mean inequality in \eqref{E:D1} one shows
\begin{equation}\label{E:D7}
\Vert{\bf u}\Vert^2\le 2R^2\Vert\theta\Vert^2+2R_s^2\Vert\phi\Vert^2.
\end{equation}
Then from \eqref{E:D6} and \eqref{E:D7},
$\Vert{\bf u}\Vert$
likewise decays as
$t\to\infty$.

\section{Conclusions}

We have extended the interesting result of \citet{Barletta:2023} for the asymptotic behaviour of the solution to convection
in a Darcy porous material for a superdiffusion model to the fully nonlinear case and we have shown that the perturbation
velocity and temperature will always decay to zero, at least in
$L^2$
norm. We have shown that this result may be extended to other convection in porous media scenarios. In particular, 
we allow for effects such as variable gravity, or convection with internal heat source. We also established an asymptotic
decay result in the important problems of bidisperse convection and double diffusive convection.

\vskip12pt

\noindent{\bf Conflict of interest}. There are no conflicts of interest.

\noindent{\bf Acknowledgement}. 
The work of BS was supported by the Leverhulme grant number EM/2019-022/9.



\end{document}